# Modeling in Systems Engineering: Conceptual Time Representation

Sabah Al-Fedaghi
*sabah.alfedaghi@ku.edu.kw, salfedaghi@yahoo.com*
Computer Engineering Department, Kuwait University, Kuwait

**Summary**
The increasing importance of such fields as embedded systems, pervasive computing, and hybrid systems control is increasing attention to the time-dependent aspects of system modeling. In this paper, we focus on modeling conceptual time. Conceptual time is time represented in conceptual modeling, where the notion of time does not always play a major role. Time modeling in computing is far from exhibiting a unified and comprehensive framework, and is often handled in an ad hoc manner. This paper contributes to the establishment of a broader understanding of time in conceptual modeling based on a software and system engineering model denoted thinging machine (TM). TM modeling is founded on a one-category ontology called a thimac (thing/machine) that is used to elaborate the design and analysis of ontological presumptions. The issue under study is a sample of abstract modeling domains as exemplified by time. The goal is to provide better understanding of the TM model by supplementing it with a conceptualization of time aspects. The results reveal new characteristics of time and related notions such as space, events, and system behavior.

## 1. Introduction

Modeling in software engineering and systems engineering involves the process of collecting and analyzing information about the system to build a model that clearly represents the domain involved. The model includes constructing static, dynamic, and behavioral representations that specify interaction among different components, as well as the architecture of the system. For example, UML 2.0 provides 14 diagram types and related modeling features and concepts that describe the system's structure as well as its behavior.

In this paper, we focus on modeling time. The increasing importance of fields such as embedded systems, pervasive computing, and hybrid systems control is bringing more attention to the time-dependent aspects of a system. The notion of time plays a major role in hardware design, computational operations, parallel processing, and real-time systems [1]. Their time aspects are concerned with temporal changes and notions such as time, behavior, event, action, state, dynamics, and concurrency. For example, UML behavioral diagrams describe the components that are dependent on time and convey the dynamics of the system. In natural language, these dynamic features are conveyed by verbs and the relationships that connect them to the passage of time.

### 1.1 Problem: Time Modeling

Yet, in computing at large, the concept of time does not always play a major role. For example, an algorithm is a process aimed at computing the value of the function; in this process, dynamic aspects are usually abstracted away [1]. Time modeling in computing is far from exhibiting a unified and comprehensive framework and is often approached in an ad hoc manner [1]. According to [2], perhaps the only characteristic common to all real-time software systems is the requirement to respond correctly to inputs within acceptable time intervals. Beyond that, the term "real-time" refers to a diverse spectrum of systems, ranging from purely time-driven to purely event-driven systems, and from soft real-time systems to hard real-time systems [2]. Accordingly, there is a need to develop an explicit conceptual time model.

### 1.2 Conceptual Modeling and Time Modeling

The scientific method involves a descriptive study or thought experiment that results in a new model (theory) whose prediction is validated by gathering data. In physics, a mathematical model of a dynamic system consists of a set of equations that state relationships between a time variable and other quantities characterizing the system [1]. Modeling refers to setting up mathematical equations to describe a system, gathering appropriate input data, and incorporating these equations and data into a computer program [3]. According to [4], "After all, a mathematical model is only a set of equations. Its link to reality is via the physical properties data of the system it is intended to simulate. A mathematical analogue can be validated only in a given number of known situations. Thus, no perfect validation is possible."

This paper focuses on a different type of modeling—conceptual modeling. Conceptual modeling is based on abstraction that represents reality (physical, social, etc.) in a simpler form. In general, conceptual modeling is concerned with identifying, analyzing, and describing the essential concepts and constraints of a domain with the help of a diagrammatic modeling language that is based on a small set of basic concepts. Conceptual modeling refers to modeling with concepts [5]. Generally, concepts may be understood according to Kant's framework of representation (see Fig. 1). Traditional thinking considers concepts as abstract meanings in their Fregean senses [6]. Concepts, as meanings, mediate between thought and language, on the one hand, and referents on the other. Each sense has a mode of presentation that represents the referent in a particular way. However, concepts are not limited to human mental representation. Concepts could exist that human beings have never entertained or may never acquire [6].

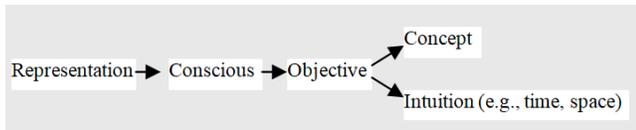

Fig. 1 Kant's view of representation (partial drawing from [7]).

### 1.3 Conceptual Time

More specifically, this paper contributes to the establishment of a broader understanding of time in conceptual modeling based on a model called thinging machine (TM). A TM model is based on a one-category ontology called a thimac (thing/machine) that is used to elaborate the design and analysis of systems' ontological presumptions. TM theory has been applied to several applications in software and systems engineering, including (to give recent publications in this journal)

- Facilitating system development processes by developing a railcar system [8]
- Further understanding UML via analyzing fine issues such as system behavior, actions, activities, etc. [9]
- Analyzing notions related to events, including Dromey's behavior trees, change over time, recurrent events, and Davidson's events [10]

This paper complements such studies by examining conceptual time. We propose a specific time foundation for the TM model. The goal is to provide better understanding of TM by supplementing it with a conceptualization of time aspects.

### 1.4 On the Nature of This Study

According to [11], scientists often navigate a tension between well-supported assertions and productive speculation reaching into uncertain territory. *Speculation* is the practice of idealizing assumptions or abstraction where "the assumption might be, or could be true" [11]. Under speculation, a hypothesis is taken as a candidate for truth introduced as part of explanatory and unifying processes [11]. The material in this paper is speculative exploration about time to complete the conceptualization of TM modeling. Hopefully, the presented materials about time do not reflect an unintended authoritative tone; rather, their origins are scattered in many sources of many great scholars, expressed in a philosophical language. Taking this into consideration, we claim the contribution in this paper is that using elaborate diagrammatic modeling (e.g., TM) is an easier way of explaining time notions (e.g., events) than most other available descriptions offer.

### 1.5 Outline of the Paper

In the next section, we briefly review related material. Section 3 introduces a new look at the ontological dual nature of things and machines in TM. Section 4 discusses TM modeling, which involves three levels: space boundaries, actionality, and dynamics. Sections 5-7 present the main contribution of the paper: conceptualizing time using TM.

## 2. Related Materials

The issue under examination is a sample of abstract modeling domains as exemplified by time. Aristotle included space and time in his ten categories and studied time order and the time point "now" [12]. Kant [12] stated, "The concept of change and, together with it, the concept of motion (as a change of the place) is possible only by and within an idea of time." Wolff [13] introduced the notion of a conceptual time system as "a pair (T,C) of two scaled many-valued contexts on the same object set G of time objects, where T is called the time part and C the event part of (T,C). The attributes in T are interpreted as time measurements, those in C as event measurements." Crang [13] discussed the idea of time existing "like beads on a string" in a sequence of isolated events, tracing its roots back to Augustine's writings is a view of the *expanded present*.

According to [14], the human conceptual system is structured around only a small set of concepts that include spatial relations, physical ontological concepts, and actions. The human conceptual system includes *conceptual time* as the time perceived in the real world or "the way events are temporally ordered with respect to each other and to the speaker" [15]. This type of time is different from linguistic time, which refers to the way time is formulated in language [15]. *Absolute time* is the time that flows uniformly without relation to anything external [16]. Time is generally conceived as a one-dimensional, directional entity. Many aspects of our concept of time are not observable in the world: Does time move horizontally or vertically? Forward or back? Left or right, up or down? Does it move past us, or do we move through it? [15].

## 3. Thinging Machines

The TM model articulates the ontology of the world in terms of an entity that is simultaneously a *thing* and a *machine*, called a *thimac* [17-20]. A thimac is like a double-sided coin. This double nature is similar to the famous Chinese yin-yang concept (see Fig. 2). As mentioned by [21], phenomena such as waterfalls, rivers, and hurricanes have dual aspects. They may be said to present both thing-like and process-like aspects [21]. Some philosophers maintain that everything we would normally take to be an object is in fact a process. In physics, particles are replaced by dynamic fields of various kinds [21]. The upper arrow in Fig. 2 denotes the thimac as a thing input into another machine. The figure includes five generic actions: create, process, release, transfer, and receive. Actions are not properties of a thing; they are machine-forms of the thimac. A thing is what can be created, processed, released, transferred, or received. The machine is a mechanism that creates, processes, releases, transfers, and/or receives things. This is an old position in science where it is said that a quality of substances (physical object) was both their acting and being acted upon (Leibniz). The machine's actions "[melt] into one another and [form] an organic whole … [the] unity thus includes a multiplicity, since it is the unity of a whole" [22]. For example, water as a machine combines the two gases oxygen and hydrogen [22].

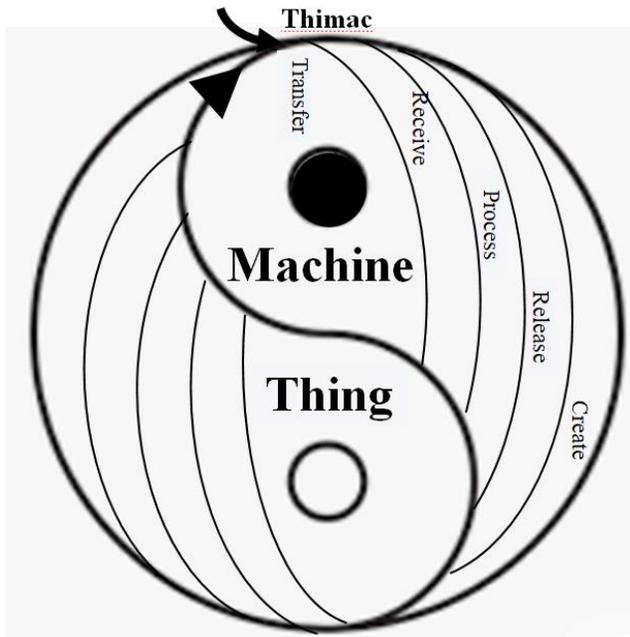

Fig. 2 As yin-yang symbol a circle divided by an S-shaped line into two segments representing a thing and machine, each is a version of the other. A thing flows into a machine and a machine becomes a thing.

The simplest type of machine is shown in Fig. 3. The actions in the machine (also called stages) are as follows:
**Arrive:**   A thing moves to another machine.
**Accept:**   A thing enters a machine. For simplification, we assume that all arriving things are accepted; hence, we can combine the thing's arrival and acceptance into the **receive** stage.
**Release:**  A thing is marked as ready for transfer outside the machine.
**Process:**  A thing is changed in form, but no new thing results.
**Create:**   A new thing is born in a machine.
**Transfer:** A thing is input into or output from a machine.

Additionally, the TM model includes storage and triggering (denoted by dashed arrows in this study's figures), which initiates a flow of things from one machine to another. Multiple machines can interact with each other through the movement of things or by triggering stages. *Triggering* is a transformation from one series of movements to another (e.g., electricity triggers creating cold air).

## 4. TM Modeling

TM modeling involves two levels (see Fig. 4), staticity and dynamics. The static model involves spatiality and actionality (to be described next). The dynamic level includes events and behavior.

4.1 Static Model: Spatiality + Actionality

The static description, denoted as S, represents the space/actionality-based description. We start static modeling by capturing activities (expressions that will be expressed in terms of the five generic actions) in reality. For example, if our target model is a pizza-ordering and delivery system, then sample observed activities would include:
- Placing an order by telephone (create)
- Paying by credit card (create, release, transfer)
- Preparing the pizza (process, create)
- Cooking the pizza (process)
- Giving the pizza to the delivery person (release, transfer)

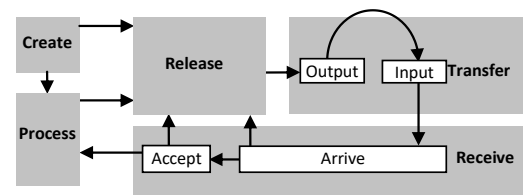

Fig. 3. The thinging machine.

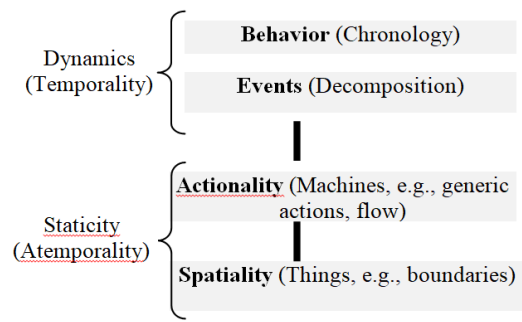

Fig. 4 TM levels

The more activities we have, the more complete the model. For example, activities involving canceling an order and refunding payment may not appear if the collected activities cover a short period.

**Example:** Spring and Hatleback [23] described the mechanism for eating a sandwich, whereby spatial things perform actions—the food, mouth, tongue, teeth, and saliva—and the actions create (e.g., saliva), process (e.g., chewing), release, transfer, and receive (e.g., moving food).

Fig. 5 shows the TM static model of this mechanism. First, the food (Circle 1) enters the mouth (2). In the mouth, the food enters the moistening stage along with the created saliva (3). In the moistening machine, the food and the saliva (4) are processed (mixed) to produce a blend. The mixture enters into the tongue's actionable sphere (5), which manipulates it to be crushed by the teeth (6).

The dynamics of the model will be developed later after discussing the static model. At this point, it is important to clarify the notion of *action* because it is a fundamental TM notion. We classify the five generic TM actions under the term *actionality* and relate them to action as used in UML, where it is claimed the activities and behavior reside.

### 4.2 Actionality Is Not Behavior

Actions in TM are related to UML actions and, more generally, UML activities. In UML, it is claimed that activity diagrams provide a high-level means of modeling dynamic system behavior [24]. Aalst et al. [24] questioned the suitability of activity diagrams for modeling processes. In UML 2.0, an action is the fundamental unit of behavior specification. "An action takes a set of inputs and converts them into a set of outputs, though either or both sets may be empty. Actions are contained in activities, which provide their context. Activities provide control and data sequencing constraints among actions as well as nested structuring mechanisms for control and scope" [25].

Traditionally, actions are called generic activities—that is, activities that cannot be divided into other activities. An activity is usually designated by a verb or verb form. Actions are identified and individuated in much the same way as entities [26]. According to [25], because an action takes input and converts it to output, action seems to be a type of PROCESS as in the well-understood input-process-output (IPO) model. The word PROCESS is capitalized to distinguish it from the action *process*, one of the generic TM actions.

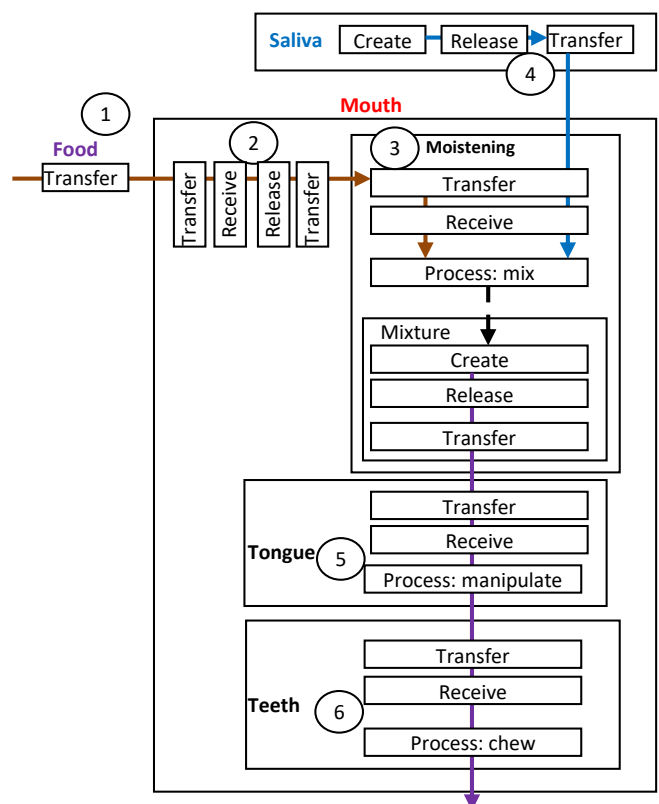

Fig. 5  Static description of the eating mechanism.

Hence, according to UML 2.0's definition and from the definitional point of view, we have the following derivation, activity → action → PROCESS. Jorgensen and Gromiec [27] identified nine types of conceptual diagrams, most of them embedding the notions of input, output, and process, either explicitly (input/output models) or implicitly (black box models). This IPO model is utilized in many interdisciplinary applications [28, 29].

What we call actionality (the five generic TM actions) is not processability, a notion stemming from PROCESS. The term PROCESS is mixed with the notions of event [21] and dynamic behavior (the UML definition mentioned previously). PROCESS is said to describe events and behavior, but such a claim is not accurate because the IPO construct does not explicitly include time. In TM modeling, actionality is a static notion that embeds the potentialities of events and behavior, which appear when time is added to the static model. In TM modeling, a thimac in the S description exists/appears in the system as a thing and as a machine, but without behavior (a time-oriented notion). S gains behavior through events. An event is formed from
- A thing, which has specific spatiality (boundary), and a machine, which has actionality; and
- Time.

In a TM model, actions at the static level are constructs signifying the mechanical form of the thimac. In the S model, the five actions do not reflect dynamism. Actionality denotes the capability (potentiality) of initiating and/or carrying out dynamism. A static thimac is the result of merging two phenomena: the spatiality of things, and actionality of machines.

4.3 Dynamic Modeling: Decomposability to Form Events and Chronology to Specify Behavior

The static model, S, only represents the steady (static) whole, so it is necessary to analyze the underlying *decompositions*, called regions, where behavior can happen (potentiality of dynamism). We can say that the static model contains all the "furniture" of the system—an inventory of everything across the past, present, and future. The furniture will be spread out when we consider the time dimension.

In imitation of the single scientific concept that recognizes the union of space and time, the TM model fuses space and time into a single dynamic model. The static description is projected as the spatiality/actionality (region) instead of a spatial coordinate system. Actually, a region is a subdiagram of S that includes spatial boundaries and actions. A union of this TM spatiality/actionality with time defines *events* as illustrated in Fig. 6, which applies the yin-yang symbol to events; the event blends such a spatiality/actionality thimac with time.

A definite thimac occupies a definite region (space +actionality) described as a subdiagram of S. Additionally, it has a definite portion of time described as its time subdiagram of events. A thimac is a construct of wider scope than *matter*. In contrast to region and time, matter/non-matter is not a fundamental conception and is represented by a subdiagram (e.g., matter box with create) of the thimac's region. The so-called motion is an eventuation of generic actions. Conservation of matter can now be thought of as the conservation of that subdiagram through events. Conservation of thimacs is an issue beyond this study.

In S (things and machines), a region reflects a conceptual space in the TM model that includes boundaries of different thimacs and static actions (see Fig. 7 for a decomposition of the mechanism for eating a sandwich). It is possible that two things may occupy the same conceptual space (i.e., a TM subdiagram), where such an arrangement is justifiable at different times. To paraphrase the famous John Wheeler, time is nature's way of keeping everything—all change that is—from happening at once. Time not only prevents all change from happening at once, but also prevents all things from existing simultaneously.

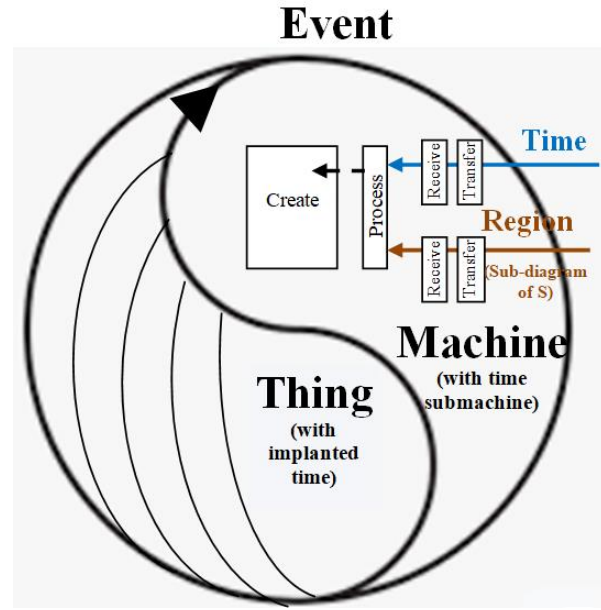

Fig. 6 The event as a machine that involves time and region

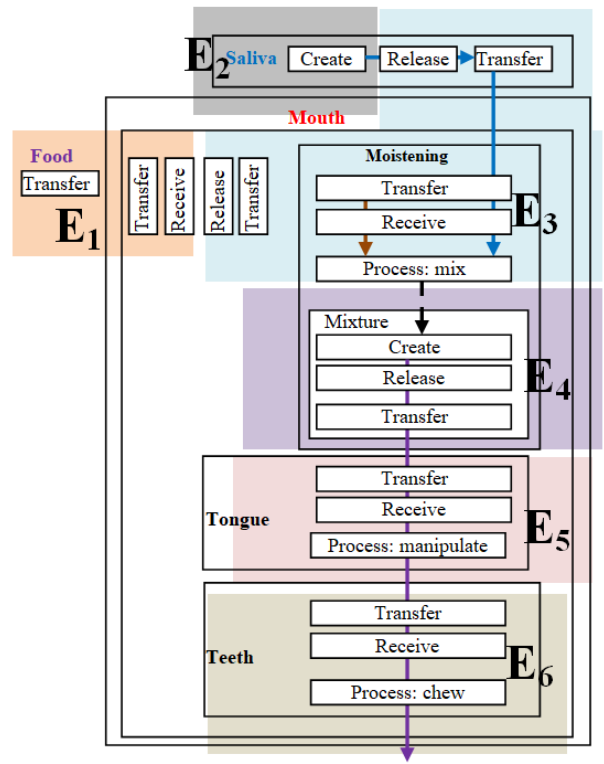

Fig. 7 Event description of the eating mechanism.

S may embed many regions of events; hence, to specify the dynamics of the system, we need to identify those portions that are susceptible to forming events when injected with time. Dividing S causes the creation of multiple subsystems, each with its own discernable spatial region. The aim of division is to achieve unity and multiplicity. Unity is reached by keeping S intact. Multiplicity is realized through the regions of S. Fig. 7 shows regions in the static model of the eating mechanism. They are labeled $E_i$s in anticipation of converting the regions to events.

The mere decomposition of S converts the system into *static areas of (potential) changes* (constitutive components) with respect to the whole S. In these static (or potential) changes (the colored areas in Fig. 7), multiplicity is a form of *becoming* from the unity (S). A static change is analogous to a set that is replaced by its members. Dynamics of a TM model here refers to decompositions of the description into areas where events occur. As mentioned, these decompositions are called regions of events. Selecting these regions is a design process.

Continuing the example of the mechanism for eating a sandwich, by injecting time, we can identify the following events (see Fig. 7).
Event 1 ($E_1$): The mouth receives the food.
Event 2 ($E_2$): The mouth generates saliva.
Event 3 ($E_3$): The mouth mixes the food and the created saliva.
Event 4 ($E_4$): A mouth generates a blend of food and saliva.
Event 5 ($E_5$): The tongue manipulates the blended matter.
Event 6 ($E_6$): The teeth crush the blended matter.
Fig. 8 shows the behavioral model of the eating machine according to the chronology of events, with the possibility of repeating some of them.

## 5. Proposed Conceptualization of TM Time

According to Theodoulidis and Loucopoulos [30], an essential issue when one is considering a model for handling the temporal dimension is the nature of the time dimension itself. In this section, we develop an explanatory frame for time in a TM. Our only criterion is to come up with speculation that seems to complement the TM model. No metaphysical issues are considered.

In a TM, we adopt the conception that time is *a spatio-temporal thimac that handles (creates, processes, releases, transfers, and receives) itself* (see Fig. 9). In TM, we adopt the conception that time is *a thimac that handles (creates, processes, releases, transfers, and receives) itself* (see Fig. 9). Thus, the thing that is transferred, received, released, processed, and created is time, and the machine that transfers, receives, releases, processes, and creates is time. This implies that time is generic (cannot be divided into other thimacs). We propose viewing time as a thing/machine for which the *now* is its current manifestation. As we saw previously in Fig. 6, time is "breathed" into a region of

S (spatiality + actionality) to generate an event. According to this view, events are just there (i.e., they then become "put down"), whereas time flows further. With this type of conceptualization, we can view *now* as the presence of time as shown in Fig. 10. *Here* is the corresponding region to *now*.

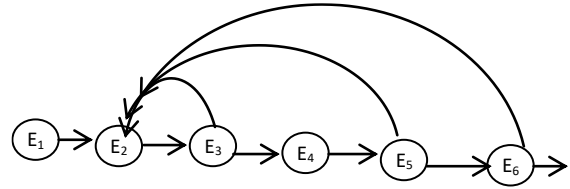

Fig. 8 Behavioral model of the eating mechanism.

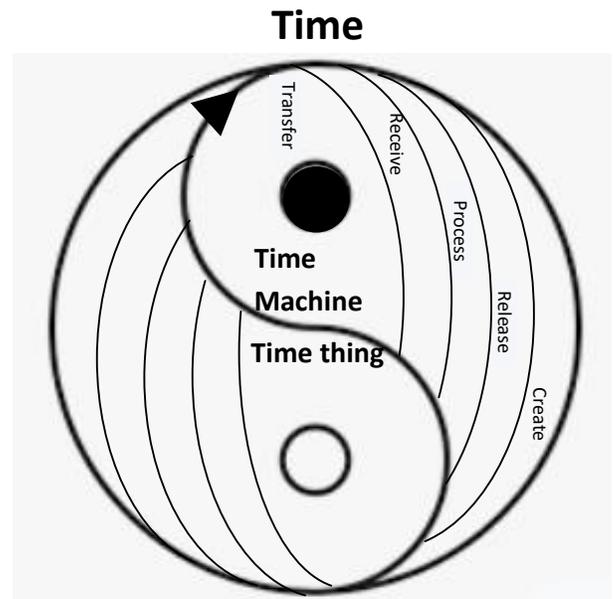

Fig. 9 The time thimac enters its machine.

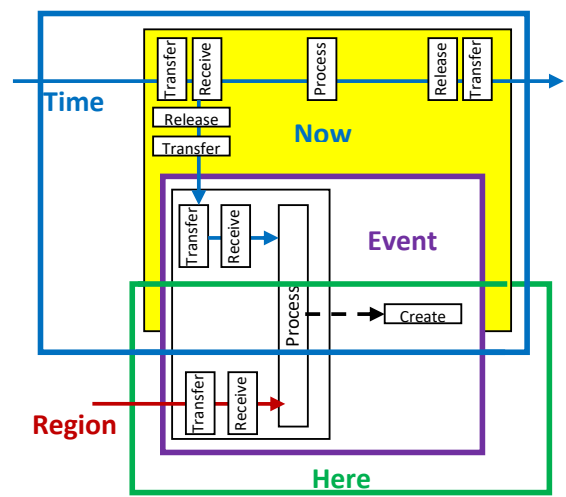

Fig. 10 The event created now.

In the figure, the generic action of the process (see the figure) mixes the time and the space (subdiagram of S) to generate the spark (triggering), which causes the eruption of the event. Here, the event is a kind of fast-living realization, just as the space (subdiagram) wakes up from a deep sleep. Accordingly, in this view, past time is merely a track littered with previous events that time has left in its forward movement. Similarly, future time is a projection of time's track. In this explanation of time in a TM, no past and no future exists, just one thing: time flowing toward somewhere. This is illustrated in Fig. 11.

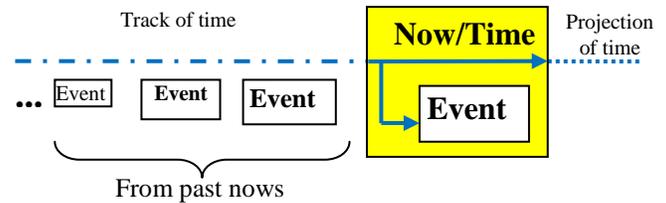

Fig. 11 Model of time

A thimac without time is a spatial/actionality thing (a pure thing, or things standardly described as being three dimensional and lacking temporality) that has no or unfunctional machine. In layman's language, such a thimac cannot act. For example, in science, it is known that matter cannot influence time or space. Thus, time "entering" thimacs is the so called nowness that we participate in and sense. Such a stand is a version of the philosophical stand called *presentism*, which adopts the philosophical position that only present things exist.

Assuming that a thing is the spatial mode of a thimac, presentism can be stated as the claim that *only events* "exist," where an event is formed from *now* and a spatial/actionality thimac. Presentism is often contrasted with two opposing views of time: eternalism (past, present, and future things exist) and, as an analogical view of modality, actualism ("only actual things exist") [32]. In the TM model, old events become static shells that exist only in records and memories (see Fig. 12) (e.g., records of bank transactions from last year or photographs). When we talk about past events, we refer to a registered record of past events in memories. The record will appear as a thing in the next *now* and will be used as a record of past events. Note that if we limit the events to generic events (based on generic actions), then the so called *light cone* will have only five possible events or less in the past, present, and future.

Furthermore, the so called *atemporal* or *timeless* (abstract) things (e.g., numbers) are things that appear (exist, create) in every *now*. Note that in a TM, "existence" means appearing in the *now* (i.e., an event), and this appearance even applies to non-moving (active) things. In a TM, to create is one of the generic actions. Integer j is a thimac that is created, processed, released, transferred, and received. An event may include a j machine and its time submachine. According to Ingram [30], presentism is consistent with the view that reality is "static" (or "frozen") and that time does not really pass.

## 6. Conceptualization to Continuity and Change

Consider the following example from Crang [13]. According to Crang [13], Augustine's "big now" is composed through the successive grasping of a future and a past. The present is always a threefold structure comprising a person's present disposition of a future and past, so the present becomes an expanded field. As an example, "a ball in flight has the just-pastness and the towards the future nearly newness of its trajectory embedded within every moment of the arc.

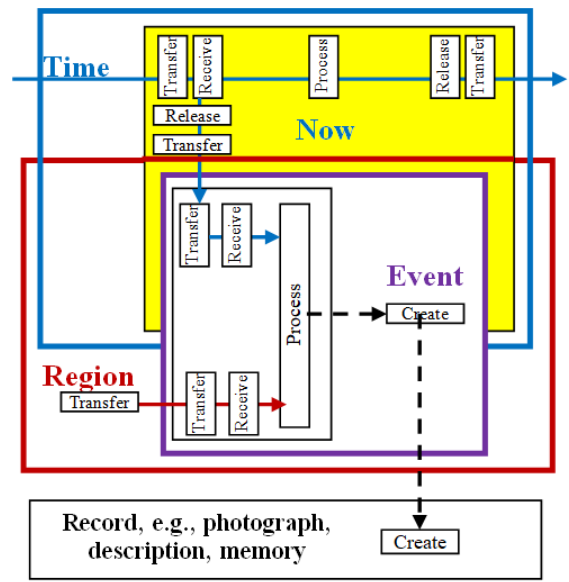

Fig. 12 The created event is registered to represent past events.

The moments are implicated, with one in the other. If this is taken as a more general pattern, it suggests that events themselves are not discrete objects or happenings but have a temporal structure" [13].

In the context of the TM model, we can model this *ball flight* example and illustrate the notions of *now* and an event. This will also illustrate the concept of space and time in a TM. Consider a ball at a specific position in an arc as shown in Fig. 13. The figure models event j, the ball crossing the spatial region, which we assume is equal to the length of the ball. Note that Fig. 13 provides multiple descriptions from the bottom up, including descriptions of the spatial region of the ball, actions, and event j. When the ball changes its position, another event j + 1 occurs, where the ball crosses a spatial region as shown in Fig. 14 (ball with dotted line). The region of event j = 1 is slightly different from the region of event j. Note that events happen to the entire ball—we cannot eventize half of the ball because the ball moves as a totality. In each position, the ball coincides with a specific space.

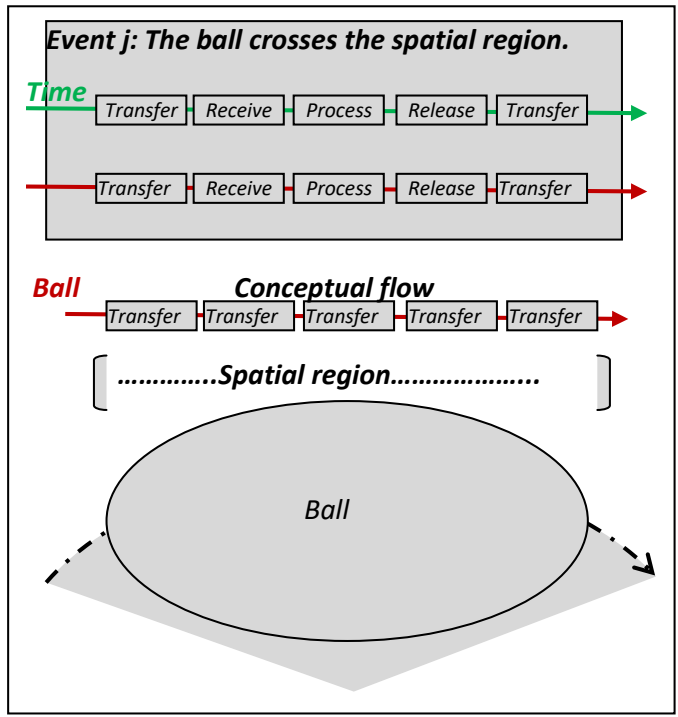

Fig. 13  The event of the ball crossing the spatial region.

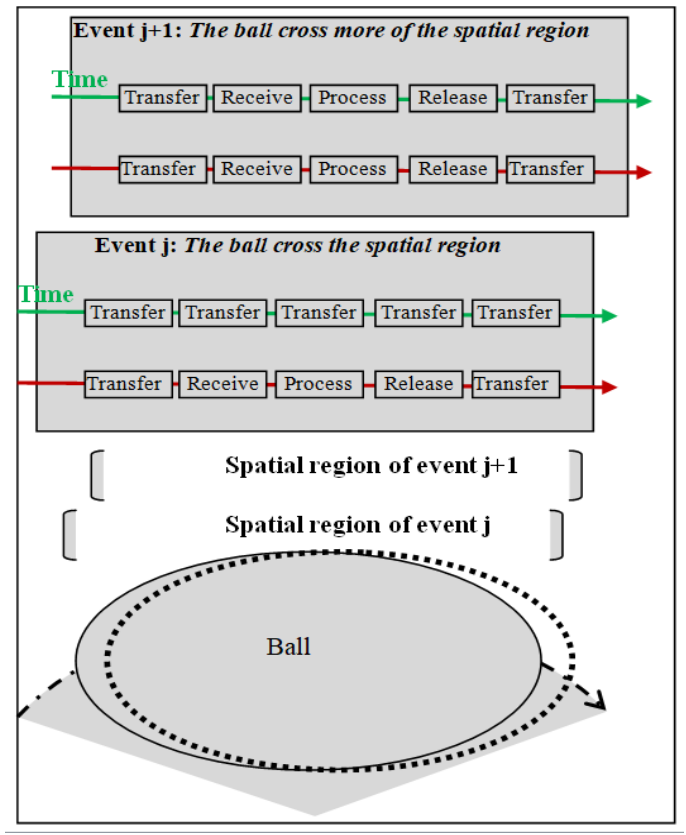

Fig. 14 The events j and j+1

We assume here a range of vision equal to the region, which can move with the ball's movement (i.e., the ball appears the same to us as a whole). The nowness (time) of the first position of j starts with transfer→receive (i.e., receiving *time* to form an event). The processing of this event, j, is cut off by the eruption of the transfer→receive of event j + 1, which generates a new nowness. The continuity through time is accomplished through eruptions of newnesses over the previous newness-s.

The *regions* of events overlap, each with its time transfer→receive depending on the ball's speed. Accordingly, we define nowness as the transfer→receive and whatever portion of the process (taking its time) is in the time sub-machines of the events. The new *now* ends with the start of the next *now*.

Time is a thing that flows continuously into overlapping regions (conceptual regions and actionality). A TM static diagram includes regions that contain actions. Time does not flow past these regions; rather, it flows continuously into them, forming their nowness. However, the time itself flows toward the future. Thus, we distinguish between time in events, and time itself flows independently of everything else.

An event is born in its *now* with its region/actions, and it stays in its (old) now until a new *now* appears. The event is a thimac made from the available spatiality/actionality and time. An event without space is time, and an event without time is space. Here, space denotes the conceptual region as represented by the TM diagram or subdiagram.

Space, in the TM sense, is the world mode of the differentiation and distinction among "beings" (thimacs). Time is the world mode of the changes in "beings." Note that no mention is made here of the relativity of time or space, which involves variability with respect to the observer (e.g., a rod in Einstein's theory has various lengths). We introduce here a view designed only for TM modeling. Our only goal is to complement TM notions, regardless of the usability of these ideas outside of a TM. The main underlying reason for such an approach is the fear of incorporating the presently huge amount of philosophical material on the subject of time, which would send the topic beyond its limited aim of proposing an (initial) foundation of time for a TM.

## 7. Case Study: Disaster-Response Scenario

Mitsch et al. [32] studied the disaster-response scenario of a gas pipe in an industrial plant leak, focusing on what would happen if an explosion occurred in an area adjacent to the pipe (see Fig. 15). In this situation, a disaster-response robot should shut off the leaking gas pipe. Partial information is provided about possible obstacles resulting from the explosion.

Debris blocks the door to the leaking gas pipe on the lower level, and a fire spreads in the area above the gas pipe. A ladder available for accessing the pipe may become inaccessible in the near future.

Fig. 16 shows the static model of this disaster-response scenario, which has been developed according to our understanding of the description by Mitsch et al. [32].

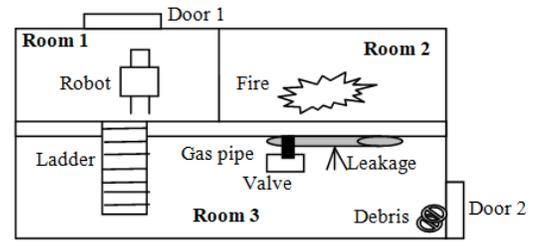

Fig. 15  Disaster-response scenario (redrawn from [32]).

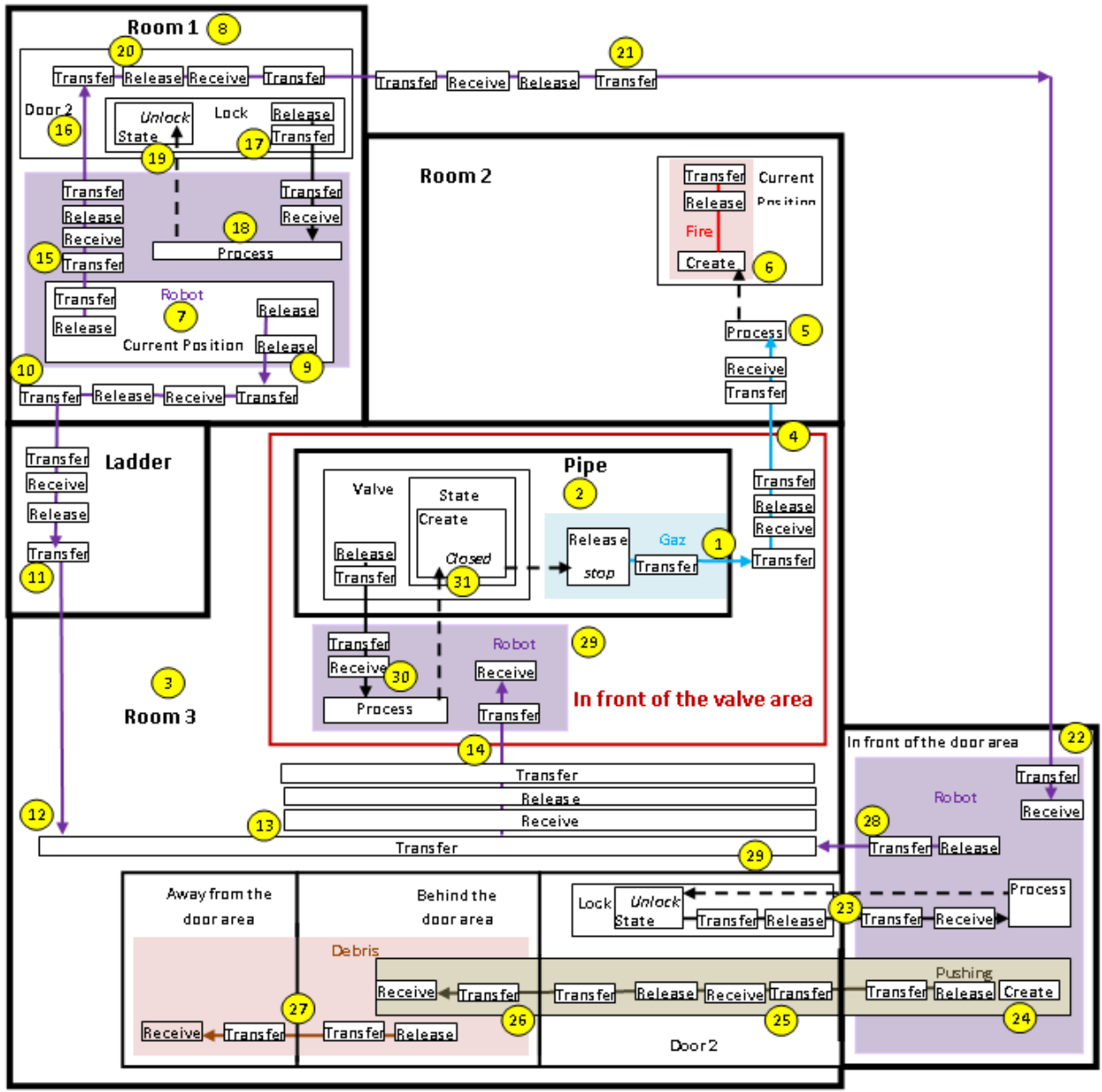

Fig. 16  The TM static model of the disaster-response scenario.

A gas leakage (circle 1) is present in the pipe (2) in room 3 (3) and has reached room 2 (4). There, it causes an explosion (5), which ignites a fire in room 2 (6). The robot in its current position (7) in room 1 (8) can reach the pipe valve that can stop the gas leakage in the following ways. Option 1: The robot moves to the ladder (9 and 10) and goes down the ladder to room 1 (11 and 12) to reach the area of the valve (13 and 14). Option 2: The robot moves to door 1 (15 and 16), where it must handle the lock (17) to process (18) and unlock it (19).

Then the robot moves to the area in front of door 2 (20, 21, and 22). There, it handles the door lock to open it (23). Additionally, the robot has to generate a pushing force to the door (24 and 25) that pushes away the debris behind the door (25), which results in moving the debris away from the door (26 and 27). Afterward, the robot enters room 3 to go to the area near the valve (28, 29, and 14).

Accordingly, we develop the dynamic model (see Fig. 17).

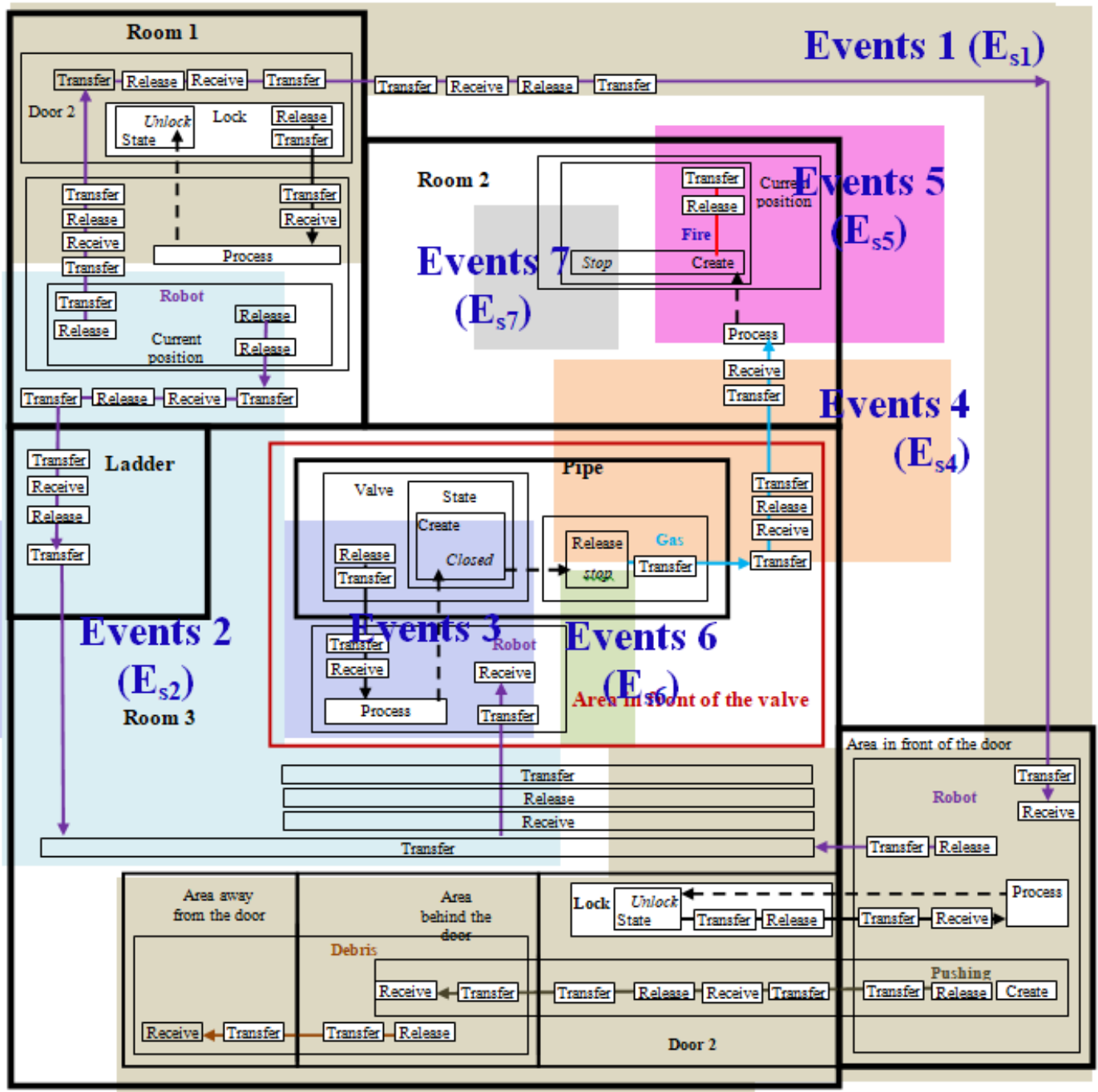

Fig. 17 The event streams in the disaster-response scenario.

Because the notion of an event has already been explained, we focus on the streams of events Es5, Es2, … , Es7 in Fig. 17, with each stream featuring consecutive sequences of events to put out the fire. Note that the robot must choose between Es1 and Es2 as its start. Fig. 18 shows the system's behavioral model. The multi circles of Es5 and Es6 denote the continuity of increased leakage as well as the growth of the fire.

Assume that the robot starts with Es2 in this case. As shown in in Fig. 19, the situation develops into a race of time between two concurrent events: the time available for the robot to stop the leakage, and the time of increased leakage and the growth of the fire until the fire becomes uncontrolled. The multi circles of Es5 and Es6 denote the continuity of increased leakage as well as the growth of the fire. The dotted lines indicate the race in the present time situation.

The fire becomes larger; hence, a new event erupts (remember the previous ball example) because the region is now different. This is illustrated in Fig. 20 for three consecutive fires. Each event is created by processing the time and region (space [the fire itself] + actionality), where both the space and the actionality extend themselves; thus, a new event is formed.

The model shows that we have successfully presented a conceptual picture of continuous time. This conceptual picture at this stage is a tool for understanding time in modeling. We can represent the picture in text, but the diagramming seems to be a more precise representation.

## 8. Conclusion

This paper contributes to establishing a broad representation of time using conceptual modeling based on the TM model. The study goal was to provide a better understanding of the TM model by supplementing it with a reasonable conceptualization of the time aspects. The results reveal new characteristics of time and related notions, such as space, events, and system behavior. A certain philosophical stand on the nature of time (presentism) is adopted because doing so seems to be suitable for TM. Accordingly, time-based analysis was added to the modeling apparatus, which complements the events and behavioral TM models.

One benefit of the paper is the apparent suitability of the TM diagrammatic method for expressing difficult notions, such as time. Future work will involve applying the methods for other philosophical approaches to time.

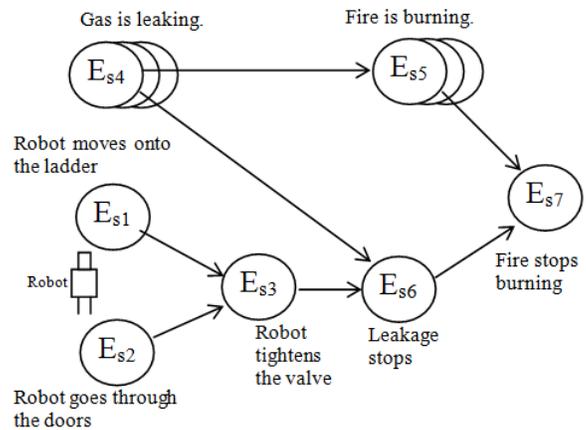

Fig. 18 The behavioral model.

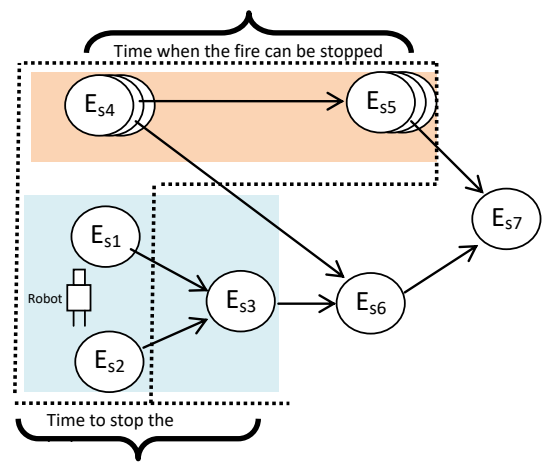

Fig. 19 The two competing events. The fire is continuously increasing in size, and the robot works to reach the valve.

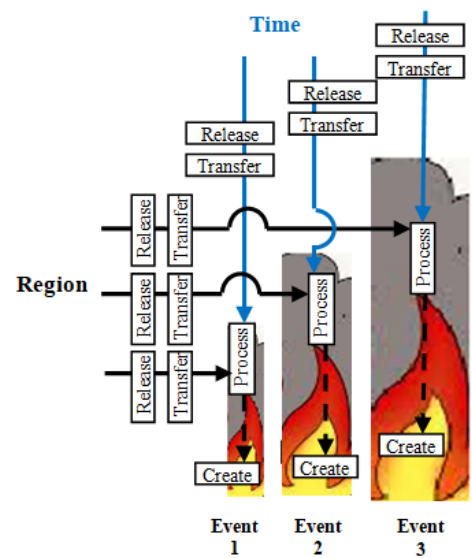

Fig. 20 Each event replaces its preceding event with differences in space/actionality.

**References**

[1] Furia, C.A., Mandrioli, D., Morzenti, A.C., Rossi, M.: *Modeling Time in Computing: A Taxonomy and a Comparative Survey*. ACM Computing Surveys 6(issue), xx–yy (2010). https://doi.org/10.1145/1667062.1667063

[2] Selic, B.: Using UML for modeling complex real-time systems. In: Mueller F., Bestavros A. (eds.) Languages, Compilers, and Tools for Embedded Systems. LCTES, Lecture Notes in Computer Science, vol. 1474. Springer, Berlin (1998). https://doi.org/10.1007/BFb0057795

[3] Kenny, I.C.: *Biomechanical and Modelling Analysis of Shaft Length Effects on Golf Driving Performance*. Faculty of Life and Health Sciences of the University of Ulster. Thesis Submitted for the Degree of Doctor of Philosophy, University of Ulster (2006)

[4] Panjabi, M.: *Validation of Mathematical Models*. Journal of Biomechanics 12(3), 238 (1979). DOI: 10.1016/0021-9290(79)90148-9

[5] Mayr, H.C., Thalheim, B.: *The Triptych of Conceptual Modeling: A Framework for a Better Understanding of Conceptual Modeling*. Software and Systems Modeling 20(issue), 7–24 (2021). https://doi.org/10.1007/s10270-020-00836-z

[6] Margolis, E., Laurence, S.: *Concepts*. In: Zalta, E.N. (ed.) The Stanford Encyclopedia of Philosophy (2021). https://plato.stanford.edu/archives/spr2021/entries/concepts/

[7] Janiak, A.: *Kant's Views on Space and Time*. In: Zalta, E.N. (ed.) The Stanford Encyclopedia of Philosophy (2020). https://plato.stanford.edu/archives/spr2020/entries/kant-spacetime/

[8] Al-Fedaghi, S.: *Diagrammatic Formalism for Complex Systems: More than One Way to Eventize a Railcar System*. International Journal of Computer Science and Network Security (IJCSNS) 21(2), 130–141 (2021). DOI: 10.22937/IJCSNS.2021.21.2.15

[9] Al-Fedaghi, S.: *UML Modeling to TM Modeling and Back*. International Journal of Computer Science and Network Security (IJCSNS) 21(1), 84–96 (2021). https://doi.org/10.22937/IJCSNS.2021.21.1.13

[10] Al-Fedaghi, S.: *Advancing Behavior Engineering: Toward Integrated Events Modeling*. International Journal of Computer Science and Network Security (IJCSNS) 20(12), 95–107 (2020). https://doi.org/10.22937/IJCSNS.2020.20.12.10

[11] Currie, A.: *Science & Speculation*. Erkenn (Preprint) 1–23 (2021). https://doi.org/10.1007/s10670-020-00370-w

[12] Wolff K.E., Yameogo W.: *Time Dimension, Objects, and Life Tracks. A Conceptual Analysis*. In: Ganter B., de Moor A., Lex W. (eds.) Conceptual Structures for Knowledge Creation and Communication. ICCS 2003. LNCS, vol. 2746, pp. 188–200. Springer, Berlin (2003). https://doi.org/10.1007/978-3-540-45091-7_13

[13] Crang, M.: *Rhythms of the City: Temporalised Space and Motion*. In: Timespace: Geographies of Temporality, pp. 187–207. Routledge, London (2001)

[14] Lakoff, G., & Johnson, M.: *Metaphors We Live By*. University of Chicago Press, Chicago (1980)

[15] Boroditsky, L.: *Metaphoric Structuring: Understanding Time through Spatial Metaphors*. Cognition 75(issue), 1–28 (2000)

[16] Wolff, K.E.: *Temporal Concept Analysis*. In: E. M. Nguifo, V. Duquenne and M. Liquière (eds.), 2001 International Workshop on Concept Lattices-Based Theory, Methods and Tools for Knowledge Discovery in Databases, pp. 91–107. Stanford University, Palo Alto, CA (2001)

[17] Al-Fedaghi, S.: *Conceptual Temporal Modeling Applied to Databases*. International Journal of Advanced Computer Science and Applications (IJACSA) 12(1), xx–yy (2021). DOI: 10.14569/IJACSA.2021.0120161

[18] Al-Fedaghi, S.: *Computer Program Decomposition and Dynamic/Behavioral Modeling*. Int. J. Comput. Sci. Netw. 20(8), 152–163 (2020). DOI: 10.22937/IJCSNS.2020.20.08.16

[19] Al-Fedaghi, S., Al-Fadhli, J.: *Thinging-Oriented Modeling of Unmanned Aerial Vehicles*. International Journal of Advanced Computer Science and Applications (IJACSA) 11(5), 610–619 (2020). DOI 10.14569/IJACSA.2020.0110575

[20] Al-Fedaghi, S., Haidar E.: *Thinging-Based Conceptual Modeling: Case Study of a Tendering System*. Journal of Computer Science 16(4), 452–466. DOI: 10.3844/jcssp.2020.452.466

[21] Galton, A.: *The Ontology of Time and Process*. Third Interdisciplinary School on Applied Ontology, Bozen-Bolzano (2016). Accessed Feb. 20, 2021. https://isao2016.inf.unibz.it/wp-content/uploads/2016/06/bolzano-notes.pdf

[22] Bergson, H.: *Time and Free Will: An Essay on the Immediate Data of Consciousness*. George Allen & Unwin, London (1950)

[23] Spring, J.M., Hatleback, E.: *Thinking About Intrusion Kill Chains as Mechanisms*. Journal of Cybersecurity 3(3), 185–197 (2017). DOI: 10.1093/cybsec/tyw012

[24] Russel, N., van der Aalst, W., ter Hofstede, A., Wohed, P.: *On the Suitability of UML 2.0 Activity Diagrams for Business Process Modelling*. In: Proc. of the Third Asia-Pacific Conference on Conceptual Modelling. APCCM (2006)

[25] OMG.: *UML 2.0 Superstructure Specification*. Technical report. In: United Modeling Language 2.0 Proposal. Sparx Systems (2004)

[26] Machamer, P., Darden, L., Craver, C.F.: *Thinking About Mechanisms*. Philos. Sci. 67(issue), 1–25 (2000)

[27] Jorgensen, S.E., Gromiec, M.J.: *Conceptual Models*. Devel. Environ. Modelling 21(issue), 211–223 (2001)

[28] Smyrk, J.R.: *The ITO Model: A Framework for Developing and Classifying Performance Indicators*. In: Australasian Evaluation Society, International Conference, Sydney, Australia (1995)

[29] Zwikael, O., Smyrk, J.: *Project Management for the Creation of Organisational Value*. In: The Input-Transform-Outcome (ITO) Model of a Project. Springer-Verlag, London (2011). DOI: 10.1007/978-1-84996-516-3_2

[30] Theodoulidis, C.I., Loucopoulos, P.: *The Time Dimension in Conceptual Modelling*. Information Systems 16(3), 273–300 (1991)

[31] Ingram, D., Tallant, J.: *Presentism*. In: Zalta, E.N. (ed.) The Stanford Encyclopedia of Philosophy (2018). https://plato.stanford.edu/archives/spr2018/entries/presentism/

[32] Mitsch, S., Platzer, A., Retschitzegger, W., Schwinger, W.: *Logic-Based Modeling Approaches for Qualitative and Hybrid Reasoning in Dynamic Spatial Systems*. ACM Computing Surveys 48(1), article 3 (2015). https:// doi.org/10.1145/2764901